\documentclass[aps,prl,showpacs,twocolumn,amsmath]{revtex4}
\usepackage{graphicx}
\def\eps{\varepsilon}

\def\E{\overline{\cal E}}

\def\k{{\bf k}}

\def\sk#1{\left(#1\right)}

\def\st#1{\left[#1\right]}

\begin{document}
\title { An improved $\eps$ expansion for three-dimensional turbulence:
summation of nearest dimensional singularities}
\author {L.Ts.~Adzhemyan$^1$, J.~Honkonen$^2$,
M.V.~Kompaniets$^1$
 and A.N.~Vasil'ev$^1$}

\affiliation{$^1$ Department of Theoretical Physics, St.~Petersburg
University, Uljanovskaja 1, St.~Petersburg,
Petrodvorez, 198504 Russia, $^2$ Theory~Division,
Department~of~Physical Sciences, P.O.~Box~64,FIN-00014
University~of~Helsinki, Finland}

\begin{abstract}
An improved $\eps$ expansion in the $d$-dimensional ($d > 2$)
stochastic theory of turbulence is constructed by taking into account pole singularities
at $d \rightarrow 2$ in coefficients of the $\eps$ expansion of universal quantities.
Effectiveness of the method  is
illustrated by a two-loop calculation of the Kolmogorov constant in three dimensions.
\end{abstract}


\maketitle

The renormalization-group (RG) method in the theory of turbulence is based on the stochastic Navier-Stokes
equation with a Gaussian random force~\cite{Nelson,Dominicis79,Adzhemyan83}. One of the central problems in
this approach is calculation of the Kolmogorov constant $C$, i.e.
the dimensionless amplitude in the scaling law~\cite{Monin}
\begin{equation}
 S_{2}(r)=C(\E r)^{2/3}  \label{S2}
\end{equation}
expressing the dependence of the second-order structure function
$S_{2}(r)$ on the relative distance $r$ in the inertial range
$r_d \ll r \ll L$. Here,  $L$ is the external length of turbulence, $r_d$ the dissipative length
and $\E$ the energy injection rate per unit mass (which, in the steady state, coincides with the
dissipation rate).

Several attempts have been
made in the past to solve this problem~\cite{UFN} - \cite{Jap}, but they all suffer from ambiguities
in connecting model parameters and observable quantities. As a consequence, there are significant discrepancies
in the predicted numerical values for the Kolmogorov constant (the spread is about a factor of two).
In this Letter we analyse reasons of this unsatisfactory situation and present results
of a calculation based both on an expression of $C$ in terms of universal quantities and account of
additional singularities arising near and below two dimensions. Rather unexpectedly, the analysis reveals
that these singularities have a major effect on the numerical values of observable quantities well above
two dimensions. We also show that a partial summation of these singularities is possible and
significantly improves the numerical value obtained for $C$.
To assess properties of the expansion produced within the RG approach, we have carried
out the calculation in the two-loop approximation (the results of Refs.~\cite{UFN} - \cite{Jap}
were obtained in the one-loop approximation).

In the RG approach to $d$-dimensional turbulence a powerlike correlation function of the random force
is often used: $\langle f\, f \rangle \sim D_0 \, k^{4-d-2\eps} \equiv d_f(k)$. In the RG framework various
quantities may be calculated in the form of an $\eps$ expansion
which subsequently must be extrapolated to the physical
value $\eps = 2$. For some important quantities the $\eps$ expansion breaks off, which for the function
$S_2(r)$ yields the Kolmogorov exponent 2/3 [as in Eq. (\ref{S2})] at $\eps = 2$.
To find the Kolmogorov constant the amplitude of this function
has to be calculated, which, however, can be
done only approximately, because its $\eps$ expansion does not break off.
In calculation of the amplitude, apart from technical difficulties at two-loop order,
a principal problem arises as well: the answer for $S_2(r)$ has to be expressed
in terms of the energy injection rate $\E$ [as in Eq.
(\ref{S2})] instead of the parameter  $D_0$ of the powerlike forcing function. This problem has been treated
in different ways in Refs. \cite{UFN} - \cite{Jap} which has led to different one-loop values of $C$.

In Ref.~\cite{JETP} (see also~\cite{UFN,turbo}) the connection between $D_0$ and  $\E$ was sought
with the aid of the
exact relation
\begin{equation}
\E = \frac{(d-1)}{2 (2\pi)^{d}} \,\int d {\bf k} \, d_f(k).
\label{bal}
\end{equation}
In the unphysical region $\eps < 2$ this integral has to be cut off at wave numbers of the order
of $\Lambda \equiv r_d^{-1}$. At fixed $\E$ this procedure introduces, first, dependence on
$\eps$ of the form $D_0 \sim (2 - \eps)$ in $D_0$ (which has to taken into account in the construction of
the $\eps$ expansion), and second, an ambiguity connected with the possibility to replace the upper limit
$\Lambda$ by $c \Lambda$  with an arbitrary coefficient $c$.
The first feature is rather natural, because the powerlike forcing
$d_f(k) \sim(2-\eps)\, k^{4-d-2\eps}$ reproduces  in the limit $\eps \rightarrow 2$
the realistic forcing by infinitely large eddies: $d_f(k) \sim\delta(\k)$.
The second feature, however, introduces arbitrariness in the sought connection
between $D_0$ and $\E$ through the coefficient $c^{2\eps-4}$, which in turn renders
the $\eps$ expansion of $D_0$-dependent quantities ambiguous
(in Ref.~\cite{JETP} the simplest choice $c = 1$ was used ). This is a reflection of
the fact that the physical content of the theory remains unaltered when
$D_0$ is multiplied by an arbitrary function $F(\eps)$ with $F(2)=1$.

Another way to fix the connection between $D_0$
and $\E$ has been used in Refs.~\cite{48} - \cite{Jap}.
It amounts to the use of an exact relation
(for the physical value $\eps = 2$ of the falloff exponent)
which allows to connect the spectral energy flux with an integral of
a third-order correlation function, the latter being subsequently calculated
in the form of an $\eps$ expansion. The use of this relation in the unphysical
region $\eps < 2$ is tantamount to a certain choice of the function
$F(\eps)$ mentioned above.

Thus, the $\eps$ expansion of the Kolmogorov constant in the model with
the powerlike forcing is not unambiguous. Therefore, a better or worse agreement
with the experimental value of $C$ at one-loop level does not bear much meaning
until a procedure for subsequent approximations has been pointed out and the
stability of obtained results checked.
On the other hand, since the real value of the expansion parameter $\eps = 2$
is not small, it is difficult to expect good quantitative results without estimating
-- at least approximately -- higher orders of the
$\eps$ expansion.

In the model at hand,
only quantities independent of $D_0$ have
rigorously unambiguous dependence on $\eps$ (we will call them universal).
Such quantities are, e.g., critical exponents and dimensionless
ratios of structure functions $S_n(r)$, the skewness factor ${\cal S} \equiv
S_{3}/S_{2}^{3/2}$ in particular. Calculation of universal quantities
with the use of the RG method and the $\eps$ expansion yields unambiguous
results and cannot lead to such ''paradoxes'' as different one-loop values for
the Kolmogorov constant.

In view of this we have pursued the goal of finding a suitable universal
quantity the physical value of which would be
simply connected with the Kolmogorov constant, and calculating this
quantity with the aid of the RG.
The skewness factor ${\cal S}$,
connected with the Kolmogorov constant through an exact relation
$C = (-4/5{\cal S} )^{2/3}$ \cite{Monin,Legacy}, might serve as such a
quantity. However, in the unphysical
region $\eps < 2$ the structure function $S_2(r)$ in the model with the powerlike
forcing correlation $d_f(k) \sim k^{4-d-2\eps}$
contains -- at $\eps < 3/2$ -- an independent of $r$ UV-divergent
additive term $\sim \Lambda^{2-4\eps/3}$ [for $S_3(r)$ this problem is
absent, see below]. As a consequence, a straightforward generalization
of the skewness factor ${\cal S} \equiv S_{3}/S_{2}^{3/2}$ to the region
$\eps < 3/2$ becomes pointless, because the powers of $r$ in this definition
do not cancel due to the constant term in $S_2(r)$. Therefore, as the desired universal
quantity we chose the ''nearest relative'' of the skewness factor, the quantity
\begin{equation}
Q(\eps)\equiv\,{  r\partial S_{2}(r)/\partial r \over |S_{3}(r)|^{2/3}}=
{   r\partial S_{2}(r)/\partial r \over (-S_{3}(r))^{2/3}} \label{Q}
\end{equation}
independent of $r$ in the whole region $0< \eps < 2$ which allows to find
the physical  values of $\cal S$ and $C$ through the simple relations
\begin{equation}
{\cal S} = - \left[\frac{2}{3Q(2)}\right]^{3/2}, \quad
C=\frac{3}{2}\st{\frac{12}{d(d+2)}}^{2/3}Q(2) .  \label{SCQ}
\end{equation}
The use of $S_3(r)$ in constructing universal quantities is advantageous
because it can be found exactly from the spectral energy balance for all $\eps < 2$:
\[
S_3(r)= -\frac{3(d-1) \, \Gamma(2-\eps) \, (r/2)^{2\eps-3}D_0}
{(4\pi)^{d/2} \, \Gamma(d/2+\eps)}\,. \label{S3}
\]
This expression allows, on one hand, to avoid calculation of graphs in construction of
the $\eps$ expansion for $S_3(r)$, and, on the other, confirms that passing to the
physical limit $\eps \rightarrow 2$, in which $\Gamma(2-\eps) \sim 1/(2-\eps)$, requires
the dependence $D_0\simeq a(2-\eps)$ to arrive at a finite value of $S_3(r)$.
The choice of the coefficient $a$ consistent with (\ref{bal})
yields the ''4/5 law'' of Kolmogorov: $S_3(r)= -\frac{4}{5} \E r$ \cite{Monin,Legacy}.

In the usual $\eps$ expansion at $d > 2$ the universal quantity $Q(\eps)$ (\ref{Q}) has the form
\cite{slovac}
\begin{equation}
Q(\eps)=\eps^{1/3}\sum_{k=0}^\infty Q_k(d)\eps^k .  \label{Qeps}
\end{equation}
The RG method allows to find successively  the coefficients of $Q_k(d)$
as a result of calculation of renormalization constants and scaling functions
in perturbation theory (loop expansion).
In Refs. \cite{UFN} -
\cite{Jap} only the one-loop approximation was used in the calculation of the Kolmogorov constant.
The results of a two-loop calculation with the aid of the relations (\ref{SCQ}), (\ref{Qeps})
have been quoted in Ref.~\cite{slovac}. For the one-loop contribution to $Q_0(d)$ in Eq. (\ref{Qeps})
an analytic expression for all $d$ may be obtained:
\begin{equation}
Q_0(d) = (1/3) [4(d+2)]^{1/3} .
 \label{Qe1}
\end{equation}
The two-loop contribution $Q_1(d)$ gives rise to integrals which may be evaluated
numerically for any particular values $d$. For $d=3$
in the calculation of the Kolmogorov constant according to Eq. (\ref{SCQ}) the values
$C^{(1)}=1.47$ (one-loop approximation)  and $C^{(2)}=3.02$
(two-loop approximation) were obtained. Although the two-loop correction
is not small, the recommended experimental value of the Kolmogorov constant $C\approx 2.0$ \cite{Monin,Sreenivasan95}
turned out to be in between the values given by the two approximations.
Hardly any more could be expected in view of the fact that the value of the expansion
parameter is not small. In what follows, we will show that the agreement with the experiment
may be significantly improved by an approximate account of the high-order terms of the
expansion (\ref{Qeps}).

Analysis of the dependence of the functions $Q_k(d)$ on the space dimension $d$ shows that
they have singularities at $d\,\leq \,2$. In particular, $Q_k(d) \sim
\Delta^{-k}$ for $2\Delta\equiv d-2 \rightarrow 0$. This means that
in the course of $d$ tending to   2 the expansion(\ref{Qeps}) necessarily will
become ''spoiled'', because the relative contribution of the high-order terms will grow
without limit. In the present two-loop approximation this feature shows in that the
ratio $Q_1(d)/Q_0(d)$ in the limit $d \rightarrow \infty$ (far away from all singularities)
is about 1/20 and monotonically grows with decreasing $d$ assuming at $d=3$ the value
$\simeq 1/2$ of which the major part is brought about by graphs singular in the limit
$2\Delta= d-2 \rightarrow 0$. This gives rise to hope that summation of leading
$\Delta$ singularities in Eq. (\ref{Qeps}) allows to improve quantitative results of the RG theory.

In the theory of turbulence the space dimension $d = 2$ is exceptional
from both the physical point of view
(additional conservation laws, inverse energy cascade) and
the formal procedure of UV renormalization, because in the limit
$d \rightarrow 2$ new divergences appear in the graphs of the
perturbation theory. These divergences show in the form of
poles in $\Delta$ in the coefficients $Q_n(d)$ for $n \geq 1$ in Eq.
(\ref{Qeps}). A consinstent procedure to remove these divergences by an additional
renormalization has been developed and gives rise to a two-parameter
$\eps, \Delta$ expansion~\cite{Nalimov}. In this Letter it is not possible
to dwell on details of calculations. However, we want to point out the following principal
issue.

The use of the $\varepsilon, \Delta$ expansion in the theory of turbulence
was proposed in Ref.~\cite{Ronis}, whose author points out that an additional
renormalization of the random force is required.
This was carried out, following Ref.~\cite{Nelson}, by multiplicative
renormalization of the random force.
There is, however, a major difference between the models of Refs.
\cite{Nelson} and \cite{Ronis}.
In Ref.~\cite{Nelson} a local correlation function of the
random force is considered ($\sim k^2$ -- model A, or $\sim const$
-- model B), wheras in Ref.~\cite{Ronis} the correlation function is nonlocal $\sim k^{4-d-2\eps}$.
From general theory of renormalization it is known that counter terms
may only be local (see, e.g.~\cite{Coll,Zinn}), which means that
the renormalization adopted in Ref.~\cite{Ronis} is inconsistent.
A consistent procedure, which we have used, was put forward in Ref.~\cite{Nalimov}.
Our two-loop calculation allowed to confirm directly the
general statement and show that it is not possible to renormalize the
present model by multiplicative renormalization of the random force.
We emphasize that, although this fact begins to show in the two-loop approximation, it
renders the corresponding one-loop result incorrect as well, since the use of the RG approach
is based on the existence of certain relations between all terms of the perturbation expansion
which break down in an incorrect renormalization. A detailed discussion of this issue we defer
to a separate publication.

In Refs.~\cite{Nalimov,Ronis} this two-parameter renormalization procedure
(in one-loop approximation) was considered
an alternative to the usual $\eps$ expansion. We exploit it in a different manner --
as a way to improve the expansion (\ref{Qeps}) by carrying out an approximate
summation of the high-order contributions.

To single out the leading poles, we express the coefficients $Q_k(d)$ in the form
\begin{equation}
Q_k(d) = \Delta^{-k}q_k(\Delta) , \, \,  2\Delta\equiv d-2 ,
 \label{Qq}
\end{equation}
with a regular function
\begin{equation}
q_k(\Delta) = \sum_{l=0}^\infty q_{kl}\, \Delta^l.
 \label{q}
\end{equation}
Substitution of the expressions from Eqs.(\ref{Qq}) and (\ref{q}) in Eq. (\ref{Qeps}) leads
for the quantity $Q$ to the representation
\begin{equation}
Q(\eps)=\eps^{1/3}\sum_{k=0}^\infty \sum_{l=0}^\infty
(\eps/\Delta)^k q_{kl} \, \Delta^l \label{Qed}\,.
\end{equation}
The $\eps,\Delta$ expansion corresponds to the asymptotic regime
$\eps \sim \Delta \rightarrow 0,\, \Delta/\eps = const$. Hence,
the quantities $(\eps/\Delta)^k$ in Eq. (\ref{Qed}) are not considered small
and the powers $\Delta^l$ play the r\^ole of a formal small parameter.
The quantity $Q$ from Eq. (\ref{Qed}) in the  $n$th-order approximation $(n\geq 1)$ is the series
\begin{equation}
 \eps^{1/3}\sum_{k=0}^\infty \sum_{l=0}^{n-1}
(\eps/\Delta)^k q_{kl} \, \Delta^l \equiv Q^{(n)}_{\eps,\Delta},
\label{Qedn}
\end{equation}
which corresponds to an approximate calculation of the
coefficients
(\ref{Qq}) of the $\eps$ expansion (\ref{Qeps}) with the account of $n$ terms in the
sum (\ref{q}). For a RG calculation of the
quantity $Q^{(n)}_{\eps,\Delta}$ in the
$\eps,\Delta$-expansion scheme~\cite{Nalimov} an $n$-loop approximation would be needed.

Let us assume for the moment that we have carried out an $n$-loop calculation
in the usual
$\eps$ expansion thus determining the following partial sum of the series (\ref{Qeps})
\begin{equation}
 \eps^{1/3}\sum_{k=0}^{n-1} Q_k(d)\eps^k \equiv Q^{(n)}_{\eps},
\label{Qen}
\end{equation}
and an $n$-loop calculation in the $\eps, \Delta$-expansion scheme as well,
hence having determined the quantity $ Q^{(n)}_{\eps,\Delta}$ of Eq. (\ref{Qedn}).
Then we have the possibility to amend the sum (\ref{Qen}) by an approximate
contribution of all higher powers of $\eps^k$ not taken into account in Eq. (\ref{Qen}).
The required information of this contribution is contained in the quantity
$Q^{(n)}_{\eps,\Delta}$. To obtain the improved value of $Q$
we add the expressions (\ref{Qedn})and (\ref{Qen}), then subtract
once the sum
\[
\delta Q^{(n)} \equiv  \eps^{1/3}\sum_{k=0}^{n-1} \sum_{l=0}^{n-1}
(\eps/\Delta)^k q_{kl} \, \Delta^l
 \label{dQen}
\]
which enters twice in the sum of Eqs. (\ref{Qedn}) and (\ref{Qen}).
Thus, we arrive at the following
$n$-loop approximation
\begin{equation}
Q^{(n)}_{eff} = Q^{(n)}_{\eps} + Q^{(n)}_{\eps,\Delta} - \delta
Q^{(n)}
 \label{Qeffn}
\end{equation}
for $Q$.
Our two-loop calculation yields the result
\begin{align}
Q^{(1)}_{\eps,\Delta}&={2}\st{\frac{2(\eps+\Delta)^2\eps}{3(2\eps+3\Delta)^2}}^{1/3}\,,\nonumber\\
{Q^{(2)}_{\eps,\Delta}\over  Q^{(1)}_{\eps,\Delta}}&=
\st{1+\sk{0.5181\eps + \frac{1}{6}\Delta}} \label{Q2ed}
\end{align}
for the quantities $Q^{(1)}_{\eps,\Delta}$ , $Q^{(2)}_{\eps,\Delta}$
with the subsequent expressions for $\delta Q^{(1)}$ , $\delta Q^{(2)}$:
\begin{align}
\delta Q^{(1)}&=\frac{2}{3}(2\eps)^{1/3}\,,\nonumber\\
{\delta Q^{(2)}\over\delta
Q^{(1)}} &=\sk{1+\frac{2\eps}{9\Delta}}\st{1+\sk{0.5181\eps +
\frac{1}{6}\Delta}}\,. \label{Qm}
\end{align}
Calculating at $d=3$ the quantity $Q^{(n)}_{\eps}$
from (\ref{Qen}) with the aid of (\ref{Qe1}) and the value
$Q_2(3)\simeq 0.4748$ found in~\cite{slovac},
and substituting the result together with Eqs.
(\ref{Q2ed}) and (\ref{Qm}) in Eq. (\ref{Qeffn}) we find the quantity
$Q_{eff}$ in first and second approximation: $Q_{eff}^{(1)}=1.38$,
$Q_{eff}^{(2)}=1.84$. Substitution of these values in Eq. (\ref{SCQ}) at
$d=3$ yields for the Kolmogorov constant and the skewness factor the values
$C_{eff}^{(1)}=1.79$,  $C_{eff}^{(2)}=2.37$,  $S_{eff}^{(1)}=-0.33$,
$S_{eff}^{(2)}=-0.22$.

\begin{table}
\caption{\label{tab:table1}
One and two-loop values of the Kolmogorov
constant in the usual $\eps$ expansion
($C_{\eps}$) and the double $\eps, \Delta$ expansion
($C_{\eps,\Delta}$); the contribution $C_{\delta}$ in Eq. (\ref{SCQ}) from the
correction $\delta Q^{(n)}$ in Eq. (\ref{Qeffn}),
and the value $C_{eff}$ from Eqs. (\ref{SCQ}), (\ref{Qeffn}).}
\begin{ruledtabular}
\begin{tabular}{lrrrr}
n&$C_{\eps}$&$C_{\eps,\Delta}$&$C_{\delta}$&$C_{eff}$\\
\hline
1 & 1.47 & 1.68 & 1.37 & 1.79 \\
2 & 3.02 & 3.57 & 4.22 & 2.37 \\
\end{tabular}
\end{ruledtabular}
\end{table}

In the Table~\ref{tab:table1} we have quoted for comparison
the values of the Kolmogorov constant calculated according to Eq. (\ref{SCQ})
in the first and second order of the usual $\eps$ expansion
($C_{\eps}$), the double $\eps, \Delta$ expansion
($C_{\eps,\Delta}$), the contribution $C_{\delta}$ in Eq. (\ref{SCQ}) from the
correction
$\delta Q^{(n)}$ in Eq. (\ref{Qeffn}) and the value $C_{eff}$ obtained from Eqs. (\ref{SCQ})
and (\ref{Qeffn}). In all the cases quoted the recommended experimental value
of the Kolmogorov constant $C\approx 2.0$ lies between the
values of the first and second approximation. However, the difference between
successive approximations is rather significant both in the
$\eps$ expansion and in the $\eps, \Delta$ expansion, let alone the
leading terms of the
$\eps$ expansion of the latter. For the improved
$\eps$ expansion, i.e. for the quantity $C_{eff}=C_{\eps}+C_{\eps,\Delta}-C_{\delta}$
calculated according to Eqs. (\ref{Qeffn}) and
(\ref{SCQ}), however, this difference is about three times smaller
leading to a far better agreement with the experimental data.

In conclusion, we have shown that a proper account of the ''nearest singularity''
in the coefficients of the $\eps$ expansion (\ref{Qeps}) leads to a
significant improvement
of the results of the two-loop RG calculation
at $d=3$. We have analysed the effect ot this procedure at other $d$ as well.
It turned out to reduce significantly the relative contribution
of the two-loop correction in the whole range considered
$\infty > d\geq 2.5$. At the same time this contribution remained large
at $d = 2$, which we think to be an effect of singularities at the next exceptional
dimension $d = 1$.

Obviously, the proposed procedure of approximate summation of the
$\varepsilon$ expansion is applicable not only to the calculation of
$Q(\eps)$, but all universal quantities such as
dimensions of composite operators.

The authors thank N.V.Antonov for
discussions. The work was supported by the Nordic Grant for
Network Cooperation with the Baltic Countries and Northwest Russia
No.~FIN-6/2002. L.Ts.A., M.V.K. and A.N.V. were also supported by
the program ``Universities of Russia''. L.Ts.A. and M.V.K.
acknowledge the Department of Physical Sciences of the University
of Helsinki for kind hospitality.

\end{document}